\documentclass[aps,twocolumn,amsmath,amssymb,prb]{revtex4-2}
\usepackage{graphicx} 
\usepackage{bm}
\usepackage{braket}
\usepackage{hyperref}
\hypersetup{colorlinks=true, urlcolor=blue, linkcolor=blue, citecolor=blue}

\begin{document}

\title{Interaction-enhanced antiferromagnetic correlations in a Kondo ladder \\
with a gapped band structure}

\author{Tatsuya Kaneko,$^1$ Ritsuki Hirabayashi,$^1$ and Yutaka Akagi$^2$}

\affiliation{$^1$Department of Physics, The University of Osaka, Toyonaka, Osaka 560-0043, Japan \\
$^2$Faculty of Core Research Natural Sciences Division, Ochanomizu University, Ohtsuka, Tokyo 112-8610, Japan}

\date{\today}

\begin{abstract}
We investigate magnetic correlations in the two-leg Kondo--Hubbard ladder using the density-matrix renormalization group method. 
We focus on a parameter regime in which the noninteracting electronic system has a narrow indirect band gap. 
In this regime, there is no Fermi surface, and the conventional Fermi-surface-based Ruderman--Kittel--Kasuya--Yosida picture is not directly applicable; magnetic correlations between localized spins become weaker as the noninteracting band gap increases. 
We find that a moderate on-site Coulomb interaction enhances the antiferromagnetic correlations between localized spins. 
This enhancement, which is also seen in the staggered magnetization induced by a pinning field at the edges, occurs for both ferromagnetic and antiferromagnetic Kondo couplings. 
Our results show that electron-electron interactions can enhance magnetic correlations between localized spins even when the underlying noninteracting electronic structure is gapped, suggesting an interaction-driven route to magnetic instabilities in narrow-gap Kondo-lattice systems.
\end{abstract}

\maketitle

\section{Introduction}

The Kondo lattice model, which describes the coupling between itinerant electrons and localized spins, is one of the fundamental models in condensed matter physics~\cite{SDoniach1977,HTsunetsugu1997}. 
In metallic Kondo-lattice systems, particularly in the weak-coupling regime, the Ruderman--Kittel--Kasuya--Yosida (RKKY) interaction mediated by itinerant electrons plays a central role in magnetic correlations between localized moments~\cite{MRuderman1954,TKasuya1956,KYosida1957}. 
The interplay between localized spins and itinerant electrons can give rise to a variety of intriguing phenomena, such as heavy-fermion behavior~\cite{PColeman2001,HLohneysen2007,PGegenwart2008,SKirchner2020} and unconventional superconductivity~\cite{EBerg2010,OBodensiek2013,MAsadzadeh2013,JOtsuki2015,PVolkov2020,JChen2024}. 
The Kondo lattice model has also been used to investigate magnetic orders with spin chirality, which give rise to topological transport phenomena~\cite{IMartin2008,YAkagi2010,YKato2010,YAkagi2012,CBatista2016,SHayami2021}. 
Understanding the properties of the Kondo lattice model therefore provides valuable insights into emergent phenomena in correlated electron systems.

A recently reported magnetic van der Waals material, GdGaI~\cite{ROkuma_arXiv,*ROkuma_arXiv_v2}, may serve as an unconventional platform for Kondo-lattice physics. 
GdGaI, which has a bilayer structure, undergoes a magnetic phase transition near 30~K, below which the localized Gd $4f$ moments are considered to form a triple-$q$ order without accompanying phonon softening~\cite{ROkuma_arXiv,*ROkuma_arXiv_v2,NJiang2026}. 
In the high-temperature paramagnetic phase, the band structure of GdGaI is semiconducting with a small indirect band gap, where the valence bands (VBs) and the conduction bands (CBs) are composed of the Gd $5d$ and Ga $4p$ orbitals~\cite{TKaneko2026}. 
Angle-resolved photoemission spectroscopy revealed that the VB maximum and the CB minimum are located at the $\Gamma$ point and the M point on the Brillouin-zone boundary, respectively, and that these bands are modified upon cooling, in association with the magnetic transition~\cite{ROkuma_arXiv,*ROkuma_arXiv_v2}. 
If the underlying band structure is gapped, there is no Fermi surface, and thus the conventional Fermi-surface-based RKKY picture is not directly applicable. 
In this regard, using the random phase approximation, one of the present authors has suggested that electron-electron interactions can enhance the magnetic instability in GdGaI~\cite{TKaneko2026}. 
This result implies that such interactions may substantially modify the coupling between localized spins in narrow-gap systems. 

\begin{figure}[b]
\begin{center}
\includegraphics[width=\columnwidth]{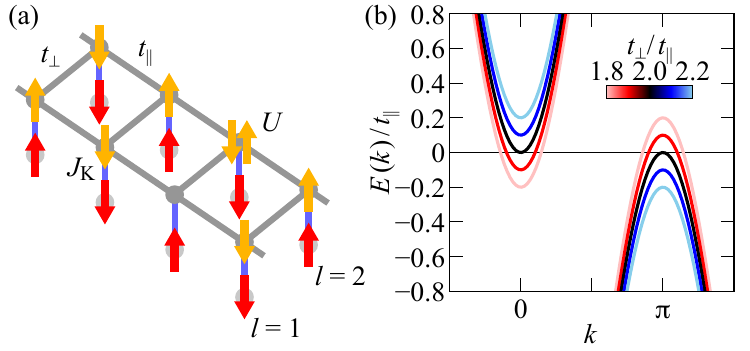}
\caption{(a) Schematic of the two-leg Kondo--Hubbard ladder at half filling. 
(b) Noninteracting ($U=0$ and $J_{\rm K}=0$) band structure of the two-leg ladder near the Fermi level for several values of $t_{\perp}/t_{\parallel}$. 
The Fermi level at half filling is set to $E_{\rm F}=0$.} 
\label{fig1}
\end{center}
\end{figure}

In this paper, we investigate the magnetic properties of the two-leg Kondo--Hubbard ladder. 
The two-leg geometry shown in Fig.~\ref{fig1}(a) provides a minimal one-dimensional setting with a tunable indirect band gap [Fig.~\ref{fig1}(b)], allowing us to study interaction effects on magnetic correlations nonperturbatively using the density-matrix renormalization group (DMRG) method. 
The calculated spin correlation function shows antiferromagnetic (AFM) correlations between localized spins at half filling, and these correlations are suppressed as the noninteracting band gap increases with $t_{\perp}$. 
However, we find that a moderate Hubbard interaction $U$ can enhance the AFM correlations. 
We also calculate the local magnetization of the localized spins in the presence of a pinning field at the edge sites and find that $U$ enhances the induced staggered magnetization, consistent with the behavior of the spin correlations. 
By examining the double occupancy and the kinetic energy of the electrons, we discuss a possible origin of this enhancement.

The rest of this paper is organized as follows. 
In Sec.~\ref{secII}, we introduce the Kondo--Hubbard model and the numerical method employed in this study. 
In Sec.~\ref{secIIIA}, we present the calculated spin correlation functions and local magnetization for ferromagnetic (FM) Kondo coupling. 
We show the results for AFM Kondo coupling in Sec.~\ref{secIIIB} and discuss the electronic properties underlying the enhancement in Sec.~\ref{secIIIC}. 
Section~\ref{secIV} summarizes our results.

\section{Model and Method} \label{secII}

We consider the Kondo--Hubbard ladder shown in Fig.~\ref{fig1}(a), in which localized spins are coupled to correlated electrons on a two-leg ladder. 
The Hamiltonian is given by
\begin{align}
\hat{H}
=&-t_{\parallel} \sum_{j,l} \sum_{\sigma} \left( \hat{c}^{\dag}_{j,l,\sigma} \hat{c}_{j+1,l,\sigma} + {\rm H.c.} \right)
\notag \\
&-t_{\perp} \sum_{j} \sum_{\sigma} \left( \hat{c}^{\dag}_{j,1,\sigma} \hat{c}_{j,2,\sigma} + {\rm H.c.} \right)
\notag \\
&+U \sum_{j,l} \hat{n}_{j,l,\uparrow} \hat{n}_{j,l,\downarrow} 
+J_{\rm K} \sum_{j,l} \hat{\bm{s}}_{j,l} \cdot \hat{\bm{S}}_{j,l}. 
\end{align}
The operator $\hat{c}^{\dag}_{j,l,\sigma}$ ($\hat{c}_{j,l,\sigma}$) creates (annihilates) an electron with spin $\sigma$~$(=\uparrow,\downarrow)$ at site $j$ on chain $l$~$(=1,2)$. 
The number and spin operators are defined as $\hat{n}_{j,l,\sigma}=\hat{c}^{\dag}_{j,l,\sigma} \hat{c}_{j,l,\sigma}$ and $\hat{\bm{s}}_{j,l}=\sum_{\alpha,\beta}\hat{c}_{j,l,\alpha}^{\dagger}\bm{\sigma}_{\alpha\beta}\hat{c}_{j,l,\beta}/2$, respectively, where $\bm{\sigma}$ is the vector of Pauli matrices. 
The operator $\hat{\bm{S}}_{j,l}$ represents an $S=1/2$ localized spin. 
The parameters $t_{\parallel}$ and $t_{\perp}$ denote the intrachain and interchain hopping integrals, respectively, and $U$ is the on-site Coulomb interaction.
The Kondo coupling between the electron spin and the localized spin on the same site is denoted by $J_{\rm K}$, where $J_{\rm K}<0$ ($J_{\rm K}>0$) corresponds to FM (AFM) coupling. 
In this paper, we assume $t_{\parallel}>0$ and $t_{\perp}>0$. 
Related Kondo--Hubbard chains have been studied, including the effects of $U$ on spin and charge excitations and on magnetic correlations~\cite{NShibata1996,AFeiguin1999}. 

In the noninteracting case ($U=0$ and $J_{\rm K}=0$), the interchain hopping $t_{\perp}$ splits the electronic bands into bonding and antibonding bands, $-2t_{\parallel}\cos k \mp t_{\perp}$, where $k$ is the wave number along the chain (the lattice constant is set to unity). 
Because the top of the bonding band and the bottom of the antibonding band are located at $k=\pi$ and $k=0$, respectively, an indirect gap $E_g=2t_{\perp}-4t_{\parallel}$ opens for $t_{\perp}/t_{\parallel}>2$. 
Figure~\ref{fig1}(b) shows the band structure near the Fermi level at half filling. 
For $t_{\perp}/t_{\parallel}<2$, both bands cross the Fermi level and the system is semimetallic, whereas for $t_{\perp}/t_{\parallel}>2$, the band structure becomes semiconducting. 
Hence, $t_{\perp}$ serves as a control parameter for the band gap in the two-leg ladder. 

For the two-leg Kondo ladder ($U=0$) at half filling with weak $J_{\rm K}$, the spin structure factor of the localized spins exhibits a peak at $\bm{q}=(q_x,q_y)=(\pi,\pi)$, corresponding to N\'eel (staggered) AFM correlations~\cite{JXavier2003,JXavier2004}. 
Here, $q_y=\pi$ represents a sign change between the two legs. 
In the semimetallic regime ($t_{\perp}/t_{\parallel}<2$), the wave vector $\bm{q}=(\pi,\pi)$ connects the Fermi points of the bonding and antibonding bands, giving rise to such AFM correlations through the RKKY mechanism at weak coupling. 
For $t_{\perp}/t_{\parallel}>2$, however, the Fermi points disappear as the noninteracting band gap opens, and this Fermi-surface-based picture is no longer applicable. 
In this study, we examine how the Hubbard interaction $U$ modifies the AFM correlations in this gapped regime.

To obtain the ground state of the Kondo--Hubbard ladder, we employ the DMRG method~\cite{SWhite1992,SWhite1993,USchollwock2011} implemented in the ITensor library~\cite{ITensor,ITensor2}. 
We consider a two-leg ladder with $L_x=64$ sites in each chain under open boundary conditions.
We focus on half filling, i.e., the system contains $2L_x$ electrons and $2L_x$ localized spins. 
We take $t_{\parallel}$ as the unit of energy. 
We keep bond dimensions up to $m= 6000$, with the largest truncation error being less than 10$^{-6}$.

\section{Results} 

\subsection{Ferromagnetic coupling ($J_{\rm K}<0$)} \label{secIIIA}

First, we consider the case of FM Kondo coupling ($J_{\rm K} < 0$). 
To examine the magnetic correlations, we compute the correlation function of the localized spins 
\begin{align}
F_{ll'}(r) = \Braket{\hat{\bm{S}}_{j_0,l} \cdot \hat{\bm{S}}_{j_0+r,l'}}, 
\end{align}
where $j_0$ is a reference site and $r$ is the distance along the chain direction. 
We set $j_0=L_x/4+1$ to reduce boundary effects. 

\begin{figure}[t]
\begin{center}
\includegraphics[width=\columnwidth]{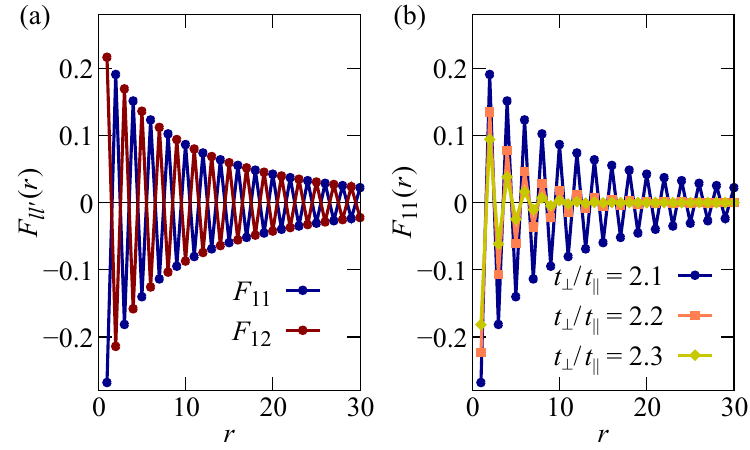}
\\
\includegraphics[width=\columnwidth]{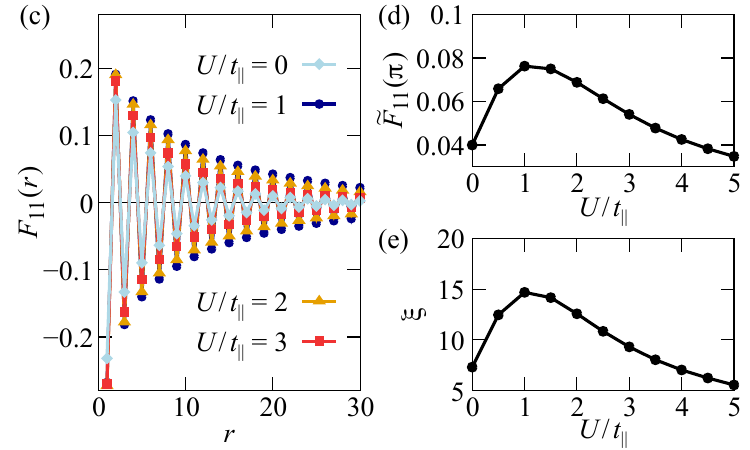}
\caption{(a) Spin correlation function $F_{ll'}(r)$ for $J_{\rm K}/t_{\parallel}=-0.5$, where $t_{\perp}/t_{\parallel}=2.1$ and $U/t_{\parallel}=1$. 
(b) $F_{11}(r)$ for various $t_{\perp}$ values, where $U/t_{\parallel}=1$. 
(c) $F_{11}(r)$ for various $U$ values, where $t_{\perp}/t_{\parallel}=2.1$. 
(d) $\widetilde{F}_{11}(\pi)$ and (e) $\xi$ as functions of $U$, where $t_{\perp}/t_{\parallel}=2.1$.} 
\label{fig2}
\end{center}
\end{figure}

Figure~\ref{fig2} shows the spin correlation function for $J_{\rm K}/t_{\parallel}=-0.5$. 
The noninteracting band structure is semiconducting at $t_{\perp}/t_{\parallel}=2.1$. 
As shown in Fig.~\ref{fig2}(a), $F_{11}(r)$ (intrachain component) and $F_{12}(r)$ (interchain component) exhibit staggered AFM correlations for $t_{\perp}/t_{\parallel}=2.1$ and $U/t_{\parallel}=1$. 
As $t_{\perp}$ increases and the noninteracting gap widens, the spin correlations are suppressed and decay more rapidly with distance [Fig.~\ref{fig2}(b)]. 
This behavior is consistent with a weakening of the electron-mediated coupling between localized spins as the noninteracting system moves deeper into the semiconducting regime.
Figure~\ref{fig2}(c) presents $F_{11}(r)$ for various values of $U$ at $t_{\perp}/t_{\parallel}=2.1$. 
The AFM correlations for $U/t_{\parallel}=1$ are stronger than those for $U/t_{\parallel}=0$. 
As $U$ increases further, however, the spin correlations are suppressed for $U/t_{\parallel}>1$. 

To quantify the $U$ dependence of the spin correlations, we define the staggered intrachain spin correlation
\begin{align}
\widetilde{F}_{ll}(\pi) = \frac{1}{r_{\rm max}} \sum^{r_{\rm max}}_{r=1} (-1)^r F_{ll}(r), 
\end{align}
where the on-site term ($r=0$) is excluded and we set $r_{\rm max}=L_x/2-1$. 
As shown in Fig.~\ref{fig2}(d), $\widetilde{F}_{11}(\pi)$ reaches a maximum around $U/t_{\parallel} \simeq 1$ and decreases as $U$ increases further.
To characterize the decay of $F_{ll}(r)$, we fit it with $F_{ll}(r)=A(-1)^re^{-r/\xi}$, where $\xi$ is the correlation length (see Appendix~\ref{appendixA} for details). 
Figure~\ref{fig2}(e) shows the $U$ dependence of $\xi$. 
As with $\widetilde{F}_{11}(\pi)$, $\xi$ is maximized around $U/t_{\parallel} \simeq 1$, indicating that the spin correlations extend to longer distances. 

\begin{figure}[t]
\begin{center}
\includegraphics[width=\columnwidth]{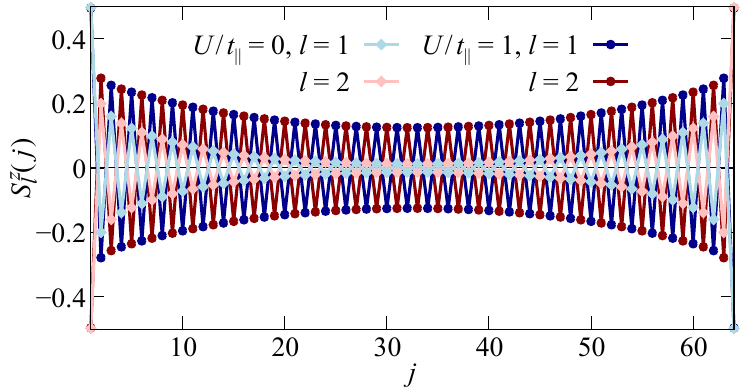}
\caption{Local magnetization $S^z_l(j)=\braket{\hat{S}^{z}_{j,l}}$ for $J_{\rm K}/t_{\parallel}=-0.5$ and $t_{\perp}/t_{\parallel}=2.1$ with a pinning field applied to the edge sites.} 
\label{fig3}
\end{center}
\end{figure}

To examine the AFM tendency in real space, we apply a pinning field to the localized spins at the edge sites, 
\begin{align}
\hat{H}_p = h_p \sum_{l} (-1)^l (\hat{S}^z_{1,l} - \hat{S}^z_{L_x,l}), 
\end{align}
and compute the local magnetization
\begin{align}
S^z_l(j) = \Braket{\hat{S}^{z}_{j,l}}. 
\end{align}
We set $h_p/t_{\parallel}=0.1$. 
Figure~\ref{fig3} shows the local magnetization $S^z_l(j)$ for $t_{\perp}/t_{\parallel}=2.1$. 
The pinning field $h_p$ polarizes the edge spins and induces local magnetization throughout the ladder. 
Consistent with the spin correlations in Fig.~\ref{fig2}(a), $S^z_l(j)$ exhibits a staggered AFM pattern, whose amplitude decays toward the center of the ladder.
As shown in Fig.~\ref{fig3}, $S^z_l(j)$ decays more slowly from the edges for $U/t_{\parallel}=1$ than for $U/t_{\parallel}=0$; the pinning-field-induced staggered magnetization penetrates farther into the ladder for $U/t_{\parallel}=1$, consistent with the enhanced spin correlations shown in Fig.~\ref{fig2}.

\begin{figure}[t]
\begin{center}
\includegraphics[width=\columnwidth]{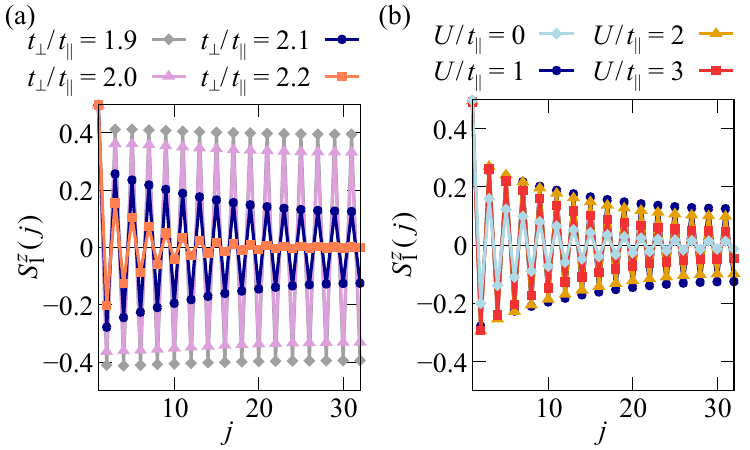}
\\
\includegraphics[width=\columnwidth]{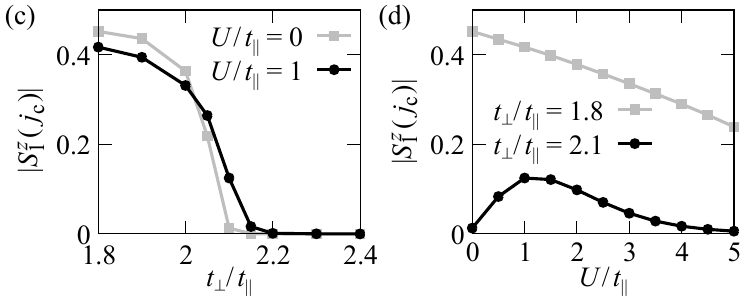}
\caption{Local magnetization $S^z_l(j)=\braket{\hat{S}^{z}_{j,l}}$ for $J_{\rm K}/t_{\parallel}=-0.5$ with a pinning field applied to the edge sites. 
(a) $S^z_1(j)$ for various $t_{\perp}$ values, where $U/t_{\parallel}=1$. 
(b) $S^z_1(j)$ for various $U$ values, where $t_{\perp}/t_{\parallel}=2.1$. 
(c) $t_{\perp}$ dependence and (d) $U$ dependence of $|S^z_1(j_{c})|$ at the center of the chain, $j_{c}=L_x/2$.} 
\label{fig4}
\end{center}
\end{figure}

Figure~\ref{fig4}(a) shows $S^z_1(j)$ for several values of $t_{\perp}$ near $t_{\perp}/t_{\parallel}=2$ at $U/t_{\parallel}=1$. 
Consistent with the behavior of $F_{11}(r)$, $S^z_1(j)$
decays steeply for $t_{\perp}/t_{\parallel}=2.2$, where the noninteracting band structure is semiconducting. 
As the gap between the bonding and antibonding bands decreases, the decay becomes slower, and the local magnetization increases near the center. 
For $t_{\perp}/t_{\parallel}<2$, where the noninteracting band structure is semimetallic, the local magnetization decays only weakly from the edges, indicating stronger electron-mediated AFM coupling. 
Figure~\ref{fig4}(b) shows $S^z_1(j)$ for various $U$ values at $t_{\perp}/t_{\parallel}=2.1$. 
Similarly to the spin correlation function, the magnitude of $S^z_1(j)$ first increases with $U$, reaches a maximum around $U/t_{\parallel} \simeq 1$, and then decreases for $U/t_{\parallel}>1$. 

To compare the cases with and without $U$, we show in Fig.~\ref{fig4}(c) the $t_{\perp}$ dependence of $|S^z_1(j_c)|$ at the center of the chain, $j_c=L_x/2$. 
As $t_{\perp}/t_{\parallel}$ increases, $|S^z_1(j_c)|$ decreases rapidly around $t_{\perp}/t_{\parallel} \simeq 2$. 
This rapid decrease is shifted to larger $t_{\perp}/t_{\parallel}$ for $U/t_{\parallel}=1$ than for $U/t_{\parallel}=0$, indicating that $U$ enhances the pinning-field-induced staggered magnetization in this parameter range. 
Figure~\ref{fig4}(d) shows the $U$ dependence of $|S^z_1(j_c)|$. 
For $t_{\perp}/t_{\parallel}=2.1$, $|S^z_1(j_c)|$ reaches a maximum around $U/t_{\parallel} \simeq 1$ and decreases as $U$ increases further, similarly to $\widetilde{F}_{11}(\pi)$ [Fig.~\ref{fig2}(d)] and $\xi$ [Fig.~\ref{fig2}(e)]. 
In contrast, for $t_{\perp}/t_{\parallel}=1.8$, $|S^z_1(j_c)|$ decreases monotonically with $U$, suggesting that the suppression of electron motion by $U$ (see Sec.~\ref{secIIIC}) weakens the correlations between localized spins.

\subsection{Antiferromagnetic coupling ($J_{\rm K}>0$)} \label{secIIIB}

\begin{figure}[t]
\begin{center}
\includegraphics[width=\columnwidth]{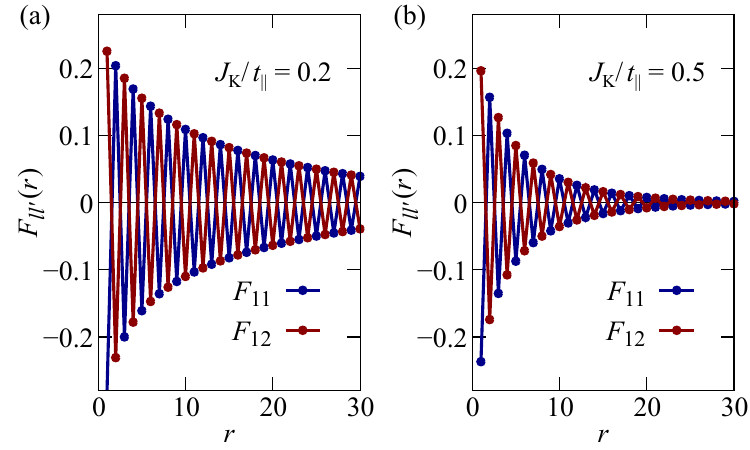}
\\
\includegraphics[width=\columnwidth]{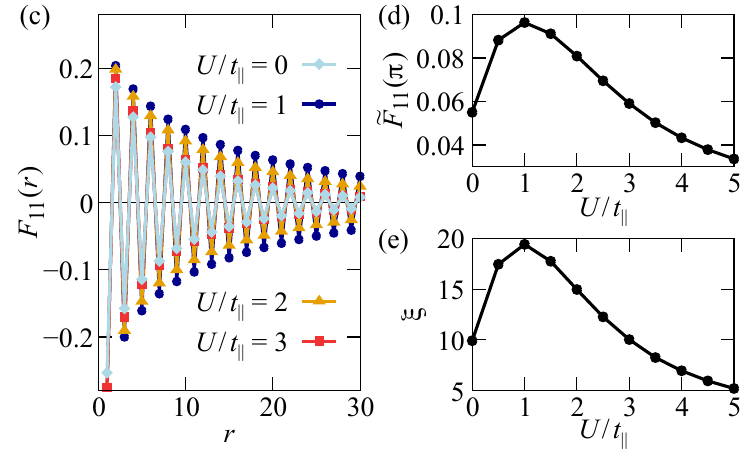}
\caption{Spin correlation function $F_{ll'}(r)$ for (a) $J_{\rm K}/t_{\parallel}=0.2$ and (b) $J_{\rm K}/t_{\parallel}=0.5$, where $t_{\perp}/t_{\parallel}=2.1$ and $U/t_{\parallel}=1$. 
(c) $F_{11}(r)$ for various $U$ values, where $J_{\rm K}/t_{\parallel}=0.2$ and $t_{\perp}/t_{\parallel}=2.1$. 
(d) $\widetilde{F}_{11}(\pi)$ and (e) $\xi$ as functions of $U$, where $J_{\rm K}/t_{\parallel}=0.2$ and $t_{\perp}/t_{\parallel}=2.1$.} 
\label{fig5}
\end{center}
\end{figure}

Next, we consider the case of AFM Kondo coupling ($J_{\rm K}>0$). 
Figures~\ref{fig5}(a) and \ref{fig5}(b) show the spin correlation function $F_{ll'}(r)$ for $J_{\rm K}/t_{\parallel}=0.2$ and $0.5$, respectively, with the other parameters the same as in Fig.~\ref{fig2}(a).
For $J_{\rm K}/t_{\parallel}=0.5$, the spin correlations are suppressed compared with those in the FM ($J_{\rm K}<0$) case with the same $|J_{\rm K}|$. 
For AFM Kondo coupling, an electron and a localized spin on the same site tend to form a spin singlet~\cite{RHirabayashi2026}, thereby screening the local magnetic moment and suppressing the AFM correlations between localized spins. 
The spin correlations become more pronounced for smaller $J_{\rm K}$, as seen by comparing Figs.~\ref{fig5}(a) and \ref{fig5}(b). 
In the following, we employ $J_{\rm K}/t_{\parallel} = 0.2$ to examine the effects of $U$. 

Figure~\ref{fig5}(c) shows $F_{11}(r)$ for various $U$ values. 
As in the FM case, the magnitude of $F_{11}(r)$ first increases up to $U/t_{\parallel} \simeq 1$ and then decreases as $U$ increases further. 
Figures~\ref{fig5}(d) and \ref{fig5}(e) show 
$\widetilde{F}_{11}(\pi)$ and $\xi$, respectively, as functions of $U$. 
Both quantities exhibit $U$ dependences similar to those for FM Kondo coupling, with maxima around $U/t_{\parallel} \simeq 1$. 
Thus, $U$ can enhance magnetic correlations for AFM Kondo coupling as well as for FM Kondo coupling.

\subsection{Electronic properties} \label{secIIIC}

\begin{figure}[t]
\begin{center}
\includegraphics[width=\columnwidth]{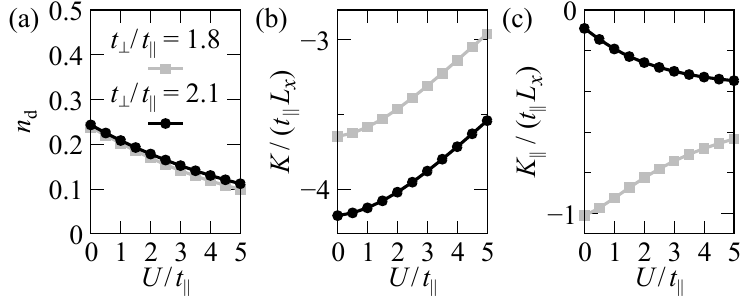}
\caption{(a) Double occupancy $n_{\rm d}$, (b) kinetic energy $K$, and (c) its chain-direction component $K_{\parallel}$ as functions of $U$ for $J_{\rm K}/t_{\parallel}=-0.5$ with a pinning field applied to the edge sites.} 
\label{fig6}
\end{center}
\end{figure}

We discuss a possible origin of the $U$-driven enhancement of the correlations between localized spins by examining how the behavior of the electrons changes with $U$. 
To characterize the localization and itinerancy of the electrons, we consider the double occupancy 
\begin{align}
n_{\rm d} = \frac{1}{2L_x}\sum_{j,l} \Braket{ \hat{n}_{j,l,\uparrow} \hat{n}_{j,l,\downarrow} }
\end{align}
and the kinetic energies associated with hopping along the chain ($\parallel$) and rung ($\perp$) directions 
\begin{align}
&\hat{K}_{\parallel}
=-t_{\parallel} \sum_{j,l} \sum_{\sigma} \left( \hat{c}^{\dag}_{j,l,\sigma} \hat{c}_{j+1,l,\sigma} + {\rm H.c.} \right), 
\\
&\hat{K}_{\perp}
=-t_{\perp} \sum_{j} \sum_{\sigma} \left( \hat{c}^{\dag}_{j,1,\sigma} \hat{c}_{j,2,\sigma} + {\rm H.c.} \right). 
\end{align}
Figures~\ref{fig6}(a) and \ref{fig6}(b) show the $U$ dependence of the double occupancy $n_{\rm d}$ and the total kinetic energy $K=\braket{\hat{K}_{\parallel}}+\braket{\hat{K}_{\perp}}$, respectively, for $t_{\perp}/t_{\parallel}=1.8$ and $2.1$, where $J_{\rm K}/t_{\parallel}=-0.5$ and $h_p/t_{\parallel}=0.1$. 
Both $n_{\rm d}$ and $|K|$ decrease monotonically with increasing $U$, consistent with enhanced localization. 
This behavior is common to both $t_{\perp}/t_{\parallel}=1.8$ and $2.1$, where the noninteracting band structure is semimetallic and semiconducting, respectively. 
Therefore, the $U$-driven enhancement of the AFM correlations for $t_{\perp}/t_{\parallel}=2.1$ cannot be explained by $n_{\rm d}$ and $K$ alone. 

A clear difference between $t_{\perp}/t_{\parallel}=1.8$ and $t_{\perp}/t_{\parallel}=2.1$ appears in the chain-direction component of the kinetic energy, $K_{\parallel}=\braket{\hat{K}_{\parallel}}$. 
As shown in Fig.~\ref{fig6}(c), $|K_{\parallel}|$ decreases with $U$ for $t_{\perp}/t_{\parallel}=1.8$, whereas it increases with $U$ for $t_{\perp}/t_{\parallel}=2.1$. 
For $t_{\perp}/t_{\parallel}=2.1$, where the noninteracting electronic system is a semiconductor, electrons form bonding states along the rungs, resulting in a small $|K_{\parallel}|$ at $U = 0$. 
As $U$ suppresses the weight of on-site doubly occupied configurations in the rung bonding states, electronic motion is redistributed from the rung direction to the chain direction, and $|K_{\parallel}|$ increases. 
This increase in $|K_{\parallel}|$ occurs in the same intermediate-$U$ regime in which the localized-spin correlations are enhanced. 
Such a coincidence suggests that the interaction-induced redistribution of electronic motion toward the chain direction may facilitate the electron-mediated coupling between localized spins. 
At larger $U$, $n_{\rm d}$ continues to decrease, while the increase in $|K_{\parallel}|$ slows down, which reflects the increasing suppression of charge fluctuations. 
The concomitant reduction of the localized-spin correlations indicates that strong electronic localization eventually counteracts the correlation-enhancing effect observed at intermediate $U$.
 
In multiorbital systems, the Coulomb interaction can shift different bands relative to each other at the Hartree level~\cite{DIhle2008,MDaghofer2010,KSeki2011}. 
In the present single-orbital ladder, however, the Hartree term shifts the bonding and antibonding bands uniformly and does not produce such a relative shift and band overlap.
For the two-leg Hubbard ladder ($J_{\rm K} = 0$), $U$ enlarges the charge gap beyond the noninteracting band gap~\cite{RMNoack1996,MKohno2019}. 
Indeed, for $t_{\perp}/t_{\parallel}=2.1$ and $J_K/t_{\parallel}=-0.5$, we also calculated the one- and two-particle charge excitation energies up to $L_x=64$ and found no indication of gap closing around $U/t_{\parallel} \simeq 1$ (not shown). 
This suggests that the enhancement of the AFM correlations around $U/t_{\parallel} \simeq 1$ is unlikely to be caused simply by an interaction-induced transition to a gapless metallic state.

\section{Summary and Outlook} \label{secIV}

We investigated the magnetic correlations in the two-leg Kondo--Hubbard ladder in the regime where the noninteracting band structure has a narrow indirect gap. 
Using DMRG, we calculated the spin correlation function and the local magnetization of the localized spins. 
We found that although the AFM correlations are suppressed as the noninteracting band gap increases with $t_{\perp}$, a moderate Hubbard interaction $U$ can enhance them. 
We confirmed the $U$-driven enhancement of the magnetic correlations for both FM and AFM Kondo couplings. 
By examining the double occupancy and the kinetic energy of the electrons, we showed that the enhancement coincides with an interaction-induced increase in the magnitude of the kinetic energy along the chain direction. 

Our study shows that electron-electron interactions can enhance magnetic correlations between localized spins even when the underlying noninteracting electronic structure is semiconducting with a narrow gap, i.e., beyond the reach of the conventional Fermi-surface-based RKKY picture.
Such an enhancement may promote magnetic ordering in higher-dimensional Kondo-lattice systems. 
This effect could be relevant to the magnetic transition in the van der Waals material GdGaI, whose band structure is considered to have a narrow gap~\cite {ROkuma_arXiv,*ROkuma_arXiv_v2}. 
In semiconductors, interband Coulomb interactions may additionally generate excitonic correlations. 
Their possible contribution to the coupling between localized spins is not directly identified from the static quantities presented here and remains an open question. 
A more direct analysis of excitonic correlations, for example through appropriate dynamical correlation functions~\cite{YMurakami2021,TKaneko2025}, would help clarify this issue. 
Regarding interaction-induced band modifications, calculations of single-particle excitation spectra~\cite{HBenthien2004,CYang2016,SPaeckel2019,TKaneko2026_2} would be useful, enabling comparison with angle-resolved photoemission experiments. 
Since Hubbard (or $t$-$J$) ladders and Kondo ladders with strong magnetic rung coupling exhibit strong superconducting correlations~\cite{EDagotto1992,MDolfi2015,TKaneko2024,HYang2025}, it is natural to investigate whether such correlations develop away from half filling in the present Kondo--Hubbard ladder and in its two-dimensional counterpart, the bilayer Kondo--Hubbard model. 
If a pronounced tendency toward superconductivity is found, designing materials that realize it would be another challenge for future work.

\begin{acknowledgments}
This work was supported by Grants-in-Aid for Scientific Research from JSPS, KAKENHI Grant No.~JP24K06939, No.~JP24H00191, No.~JP24K01333, and No.~JP24K00546, and JST PRESTO, Japan, Grant No.~JPMJPR2251. 
The DMRG calculations were performed using the ITensor library~\cite{ITensor,ITensor2}.
\end{acknowledgments}

\appendix 

\section{Fitting of the spin correlation function} \label{appendixA}

\begin{figure}[b]
\begin{center}
\includegraphics[width=\columnwidth]{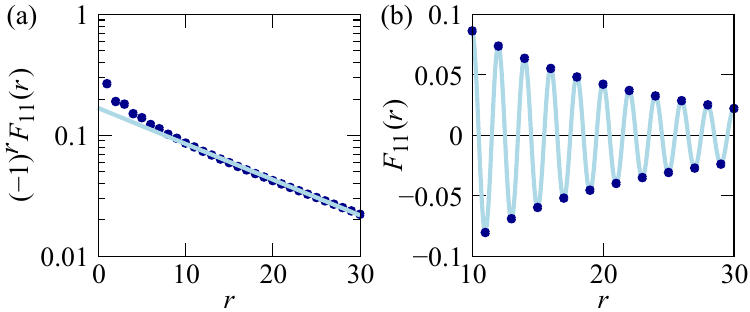}
\caption{Fitting of the spin correlation function $F_{11}(r)$, where $J_{\rm K}/t_{\parallel}=-0.5$, $t_{\perp}/t_{\parallel}=2.1$, and $U/t_{\parallel}=1$. 
(a) Semilogarithmic plot of $(-1)^rF_{11}(r)$ and (b) linear-scale plot of $F_{11}(r)$, where the solid lines represent the fitting curves.} 
\label{fig7}
\end{center}
\end{figure}

Since the fitting function $F_{ll}(r)=A(-1)^re^{-r/\xi}$ decays exponentially, we fit $\ln|F_{ll}(r)|$ with the linear function $-r/\xi+\ln A$. 
To capture the long-distance decay, we use the data for $8 \le r \le L_x/2$ in the fitting. 
Figure~\ref{fig7}(a) shows $(-1)^rF_{11}(r)$ and the fitting curve on a semilogarithmic scale. 
The values of $\xi$ shown in Figs.~\ref{fig2}(e) and \ref{fig5}(e) are obtained by this procedure. 
As shown in Fig.~\ref{fig7}(b), the fitting function reproduces the spin correlations well at long distances ($r \gtrsim 10$). 

\bibliography{references}

\end{document}